\documentclass{elsart}
\usepackage{graphics}

\begin{document}
\begin{frontmatter}
\title{Studies of 100~\mbox{\boldmath $\mu$}m-thick silicon strip 
detector with analog VLSI readout}

\author[RCNP]{T. Hotta\thanksref{contact}},
\author[RCNP]{M. Fujiwara},
\author[Yamagata]{T. Kinashi},
\author[KEK]{Y. Kuno},
\author[RCNP]{M. Kuss\thanksref{cebaf}},
\author[RCNP]{T. Matsumura},
\author[RCNP]{T. Nakano},
\author[Tsukuba]{S. Sekikawa},
\author[Tokyo]{H. Tajima}
and
\author[RCNP]{K. Takanashi}
\address[RCNP]{Research Center for Nuclear Physics, Osaka University,
Ibaraki, Osaka 567, Japan}
\address[Yamagata]{Department of Physics, Yamagata University Yamagata,
Yamagata 990, Japan}
\address[KEK]{Institute of Particle and Nuclear Studies (IPNS),
High Energy Accelerator Research Organization (KEK), Tsukuba, Ibaraki 305,
Japan}
\address[Tsukuba]{Institute of Physics, University of Tsukuba, Tsukuba,
Ibaraki 305, Japan}
\address[Tokyo]{Department of Physics, University of Tokyo, Bunkyo-ku,
Tokyo 113, Japan}
\thanks[contact]{Corresponding author. e-mail hotta@rcnp.osaka-u.ac.jp}
\thanks[cebaf]{Present Address: Jefferson Laboratory, Newport News, VA 23606,
USA}
\begin{abstract}

We evaluate the
performances of a 100~$\mu$m-thick silicon strip detector (SSD) 
with a 300 MeV proton beam and a $^{90}$Sr 
$\beta$-ray source.  
Signals from the SSD have been read out using a VLSI chip.
Common-mode noise, signal separation efficiency and energy
resolution are compared with those for the SSD's with a thickness 
of 300~$\mu$m and 500~$\mu$m.
Energy resolution for minimum ionizing particles (MIP's) 
is improved by fitting the non-constant component in a 
common-mode noise with a linear function.

\end{abstract}
\end{frontmatter}
\section{Introduction}

A silicon strip detector (SSD) has the highest
position resolution among the electric tracking devices in
particle physics experiments. 
However, an error in measuring the track angle 
is dominated by the multiple scattering effect for 
particles with a low velocity.
If the effect is reduced with a very thin SSD, 
new experiments which are impossible by the present technology
will be realized.
One example is a search for the $T$ violation
in the decay of B mesons~\cite{kuno},
in which the $T$-violating transverse $\tau^{+}$ polarization
in the decay ${\mathrm B} \to {\mathrm D}\tau^{+} \nu$
will be measured to a precision of 10$^{-2}$.
In order to obtain the $\tau$ polarization
the decay vertices of B and $\tau$ must be measured separately.
A simulation shows that the experiment will be feasible only 
with very thin SSD's at asymmetric-energy B factories.

In general, a thin SSD has a small signal-to-noise (S/N)
ratio because the energy deposit in the detector is proportional
to the thickness 
and its large capacitance results in a large noise.
Thus careful treatment of a noise in the off-line analysis is
important.

In this paper, we evaluate the performances of a 100~$\mu$m-thick
silicon strip 
detector.
The performances are compared with those of the 300~$\mu$m and 
500~$\mu$m-thick silicon strip detectors.

\section{Detector}

Single-sided silicon detectors 
with the dimensions of 1 cm $\times$ 1.3 cm have been fabricated by
Hamamatsu Photonics.  
The strip pitch is 100~$\mu$m.
The widths of implantation strips and aluminum electrodes are 42~$\mu$m
and 34~$\mu$m, respectively.
Three detectors with different thicknesses (100~$\mu$m, 300~$\mu$m, and 
500~$\mu$m) were tested.
The 100~$\mu$m-thick SSD was made by etching a 300~$\mu$m-thick wafer.
The analog VLSI chips (VA2\footnote{Produced by Integrated Detector
and Electronics AS (IDEAS), Oslo, Norway.})~\cite{VA2} are used as a readout
circuit of the detectors. 
An SSD and a VLSI chip were mounted on a printed circuit board
called ``hybrid board''.

\section{Experiment}

Two different particles were used for 
evaluation of the detector performances.
A proton beam was used to measure the response of detectors
for baryons or heavy particles.
To see the response for light and high velocity particles which
satisfy the minimum ionizing particle (MIP) condition ($E/m > 3$),
electrons from a \nuc{90}{Sr} $\beta$-ray source was used.

The experiment was carried out with a proton beam at the 
Research Center for Nuclear Physics, Osaka University.
Scattered protons from a \nuc{12}{C} target were momentum analyzed by a
magnetic spectrometer.
A detector system that consists of an SSD and two trigger plastic
scintillation counters was placed at the focal plane of the spectrometer.
The momentum of detected  protons was 800~MeV/$c$ with 
the momentum spread of $< 0.05 \%$.
The energy loss for a proton with 800~MeV/$c$ 
is 68 keV for the 100~$\mu$m-thick SSD, which is about 
1.7 times larger than that for the minimum ionizing protons.

\begin{figure}
\begin{center}
\includegraphics{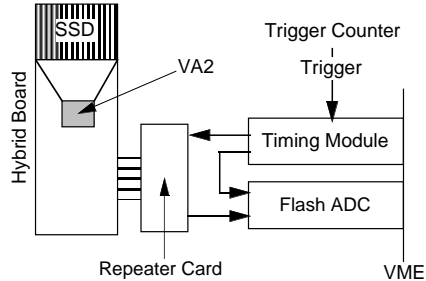}
\end{center}
\caption{The schematic view of the readout system.}
\label{fig:readout}
\end{figure}

The readout system is schematically shown in Fig.~\ref{fig:readout}.
The hybrid board consisting of a silicon strip and a VA2 chip
was connected to a ``repeater card'',
which contained
level converters for logic signals, 
buffer amplifier for analog output signal,
and adjustable bias supply for the VA2 chip. 
The VA2 chip was controlled by a VME based timing board which received
a trigger signal and generated clock pulses for VA2 and
a VME based flash ADC board.  
Analog multiplexed output from VA2 was sent to a flash ADC 
through the repeater card.
Two layers of trigger counters were placed in
front of the SSD.
The repeater card was connected to the hybrid board with
a ribbon cable for both the analog and logic signals. 
The length of the ribbon cable was about 15~cm.

In order to compare the characteristics of silicon strip detectors,
the operation parameters of the VA2 readout chips were 
fixed to standard values without optimization for each measurement.
Signal shaping time was about 700 ns.
Signals were read out in 4 $\mu$sec clock repetition.
Typical trigger rate was about 30~Hz.

In addition to a proton beam test, 
measurements with a \nuc{90}{Sr} $\beta$-ray source were also 
performed.
The \nuc{90}{Sr} $\beta$-ray source was placed at 15~mm
from the SSD.
A collimator with a size of 2~mm in diameter and 10~mm in thickness was
used to irradiate electrons perpendicularly to the SSD.
In order to realize the minimum ionizing condition,
a high energy component of $\beta$-rays was selected 
by a
trigger scintillation counter placed behind the SSD.
Operation parameters of the VA2 chip was the same as those at the proton
beam test.
Readout clock was 400 ns.
The trigger rate at the $\beta$-ray source test was about 7~Hz.

\section{Analysis and Results}

\begin{figure}
\begin{center}
\includegraphics{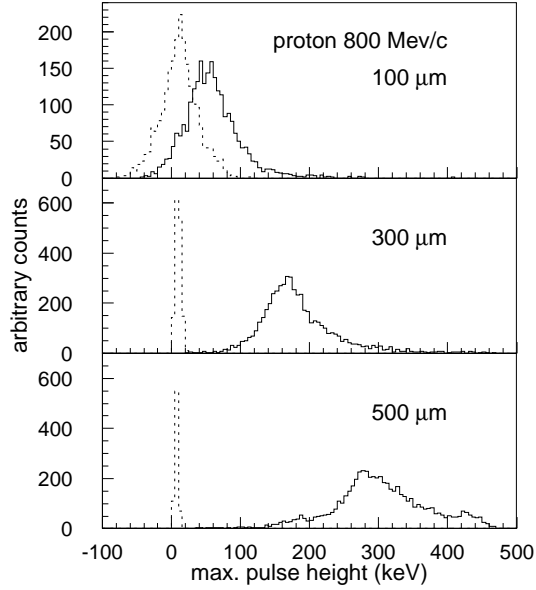}
\end{center}
\caption{The maximum pulse height for proton signals (solid lines) 
and noises (dotted lines) for the 100~$\mu$m, 300~$\mu$m, and
500~$\mu$m-thick SSD's.}
\label{fig:proton-bef-cms}
\end{figure}

An output from each strip has a different offset level.
These differences have been trimmed at the first step of the off-line analysis.
Solid lines in Fig.~\ref{fig:proton-bef-cms} show the
maximum pulse height distributions after the pedestal trimming 
for 100~$\mu$m, 300~$\mu$m, and
500~$\mu$m-thick SSD's at the proton beam test.
Note that we have neglected the effect of charge division among
adjacent strip.
Dotted lines show the same distributions under the condition
that no charged particle hit the detector.
The noise peak and proton signal peak have overlapped for the 
100~$\mu$m-thick SSD, 
while the proton signals are clearly distinguished from noises 
for the 500~$\mu$m and 300~$\mu$m-thick SSD's.

\begin{figure}
\begin{center}
\includegraphics{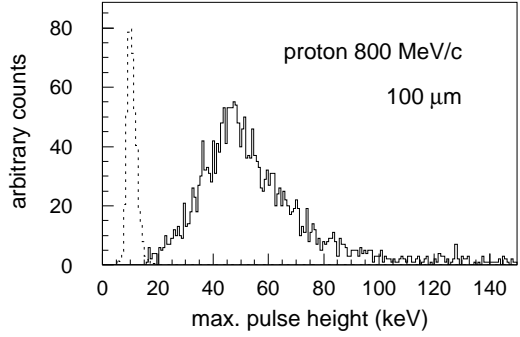}
\end{center}
\caption{Maximum pulse height for protons (solid line) and noise (dotted line)
after the CMN subtraction for the 100~$\mu$m-thick SSD. }
\label{fig:proton-aft-cms}
\end{figure}

For 100~$\mu$m-thick SSD, a strong noise level correlation between
non-adjacent channels has been observed.
This indicates that the main component of the noise has a common phase
and amplitude among the strips.
This component called common-mode noise (CMN) 
has been calculated as an averaged pulse 
height over all strips.
In the calculation, channels with significantly large 
pulse heights; larger than 3 standard deviation $(\sigma)$ 
of the noise distribution, have been excluded.
Fig.~\ref{fig:proton-aft-cms} shows
the maximum pulse height distribution
after the CMN subtraction
for the 100~$\mu$m SSD.
Proton events are clearly separated from the noise.

\begin{figure}
\begin{center}
\includegraphics{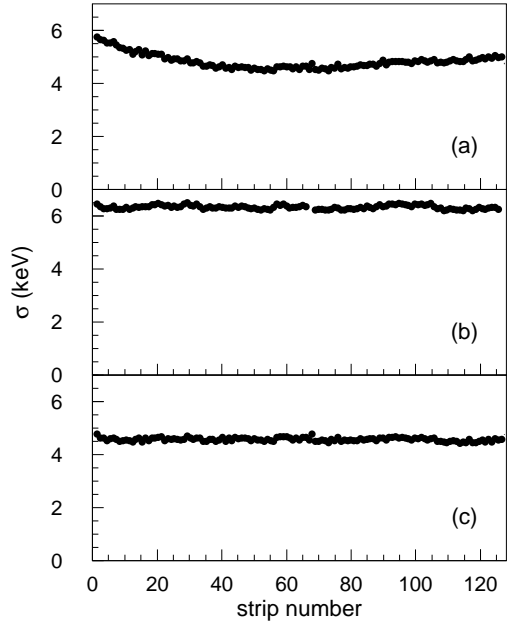}
\end{center}
\caption{Strip dependence of the noise width, $\sigma$.
(a) After subtracting the constant CMN, (b) width of the difference between 
adjacent strips, (c) After the CMN subtraction by linear-fitting.}
\label{fig:noise-char}
\end{figure}

We have investigated the characteristics of noise more carefully.
Fig.~\ref{fig:noise-char}(a) shows the strip dependence of the noise
width after the CMN subtraction.
The width depends on the strip number,
whereas pulse height differences between adjacent two
strips shown in Fig.~\ref{fig:noise-char}(b) have a constant value 
of about 6.
This indicates that the intrinsic $\sigma$ of the noise
is expected to be about 4.2 ($=6/\sqrt{2}$)
for all strips. 
Thus, we conclude that the CMN has a 
non-constant component.
Instead of simply averaging the pulse heights,
we fit them with a linear curve to get CMN as a function of a channel
number.
Fig.~\ref{fig:noise-char}(c) shows the noise widths after this 
method is applied.
The widths are about 4.2 for all strips as expected.

\begin{figure}
\begin{center}
\includegraphics{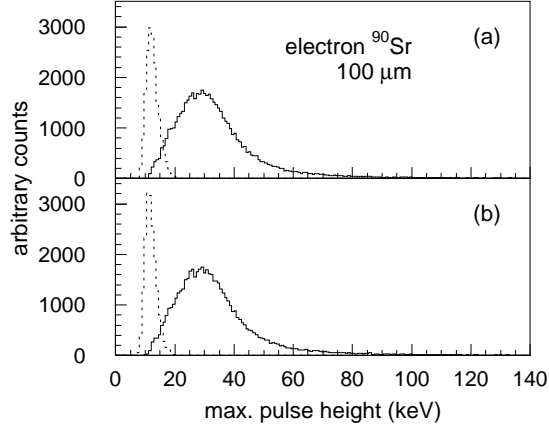}
\end{center}
\caption{Maximum pulse height of electrons from \nuc{90}{Sr} source 
for the 100~$\mu$m SSD after the CMN subtraction by constant (a) and 
linear-fitting (b).}
\label{fig:electrons}
\end{figure}

If the CMN is not removed correctly by assuming a constant CMN,
a noise width depends on a strip number (Fig.~\ref{fig:noise-char}(a)).
This may cause a strip dependent S/N separation which are not desirable
for any experiments.
Fitting the CMN with a linear curve 
is particularly important for the detection of MIP's with a thin SSD 
where the S/N ratio is small.
The maximum pulse height distribution
for electrons with 100~$\mu$m-thick SSD after subtracting the CMN by
linear-fitting is shown in Fig.~\ref{fig:electrons}(b) compared with that with
a constant CMN subtraction (Fig.~\ref{fig:electrons}(a)).
Although electron events are not separated from the noise for both cases,
the separation of signals from noises is improved by the linear-fitting
method\footnote{$\beta$-rays were irradiated at the central strips by using
a collimator. It is expected that this improvement is clearly seen for
the strips near the edge of the detector.}.

\begin{figure}
\begin{center}
\includegraphics{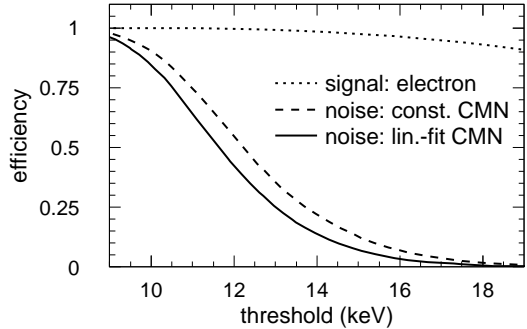}
\end{center}
\caption{Detection efficiency for electrons (dotted curve) 
with the 100~$\mu$m-thick SSD. Solid and dashed curves indicate the
fraction of noise peak after the constant CMN-subtraction 
and the linear-fitting methods were applied, respectively.}
\label{fig:efficiency}
\end{figure}

Fig.~\ref{fig:electrons} indicates that
there is a finite probability of misidentifying 
a noise as a particle track by selecting the maximum pulse height.
The detection efficiency and signal misidentification probability 
for electrons with 100~$\mu$m-thick SSD
are plotted  as a function of threshold energies in Fig.~\ref{fig:efficiency}.
When a threshold level is 
set to detect the electron with an efficiency more than 99\%
the probability of misidentification
obtained by linear-fitting of CMN is 27\% smaller 
compared to that by the constant CMN-subtraction method.
  
The S/N ratio obtained from $\beta$-ray source tests for 
the 100~$\mu$m and 300~$\mu$m SSD's
are summarized in Table~\ref{tab:eleSN}.
\begin{table}
\caption{S/N ratios for $\beta$-ray electron signal.}
\begin{tabular}{lrr} \hline
SSD thickness & 100~$\mu$m &  300~$\mu$m  \\ \hline 
without CMN subtraction & 4.91  & 17.1  \\
constant CMN & 7.45 & 28.7  \\
linear-fitted CMN  & 7.88 & 29.7  \\ \hline 
\end{tabular}
\label{tab:eleSN}
\end{table}
Better S/N ratio is obtained by fitting CMN with a linear
curve.
The S/N ratios obtained with the assumption of a constant CMN for both 
the 100~$\mu$m and
300~$\mu$m SSD's are slightly worse.
For a 300~$\mu$m SSD,
the difference of two methods in subtracting the CMN is 
not very important in an actual application
because the S/N ratio is sufficiently large.
Noise width obtained in the
\nuc{90}{Sr} $\beta$-ray
source test and the proton beam test 
are summarized in Table~\ref{tab:noise-kev} in energy unit (keV).
\begin{table}
\caption{Noise width[keV] at the \nuc{90}{Sr} $\beta$-ray source test
(and proton beam test).}
\begin{tabular}{lrrrrrr} \hline
SSD thickness & \multicolumn{2}{c}{100~$\mu$m} & \multicolumn{2}{c}{300~$\mu$m}
& \multicolumn{2}{c}{500~$\mu$m} \\ \hline 
no CMN subtraction & 7.34 & (27.7) & 6.27 & (4.77) & --- & (3.27)  \\
constant CMN & 4.83 & (4.18) & 3.73 & (3.58) & --- & (2.89)  \\
linear-fitted CMN & 4.57 & (4.14) & 3.60 & (3.56) & --- & (2.84) \\ \hline 
\end{tabular}
\label{tab:noise-kev}
\end{table}
The width of CMN at the $\beta$-ray source test is different from
that at the proton beam test.
But the noise after the CMN subtraction is almost the same.

There remains a possibility to improve the S/N ratio by
considering the charge division among adjacent strips during
finding a particle trajectory.
Performances of the prototype detector might be
improved by optimizing its operating conditions.

\section{Conclusion}
An SSD with a thickness of 100~$\mu$m was tested with 800 MeV/$c$ protons 
and $\beta$-rays from \nuc{90}{Sr} source.
By using an analog VLSI chip for readout, we remove the CMN. 
Assuming that CMN is constant among all strips, 
proton signals are separated from noises for the 100~$\mu$m, 300~$\mu$m
and 500~$\mu$m-thick SSD's after the CMN subtraction.
We found that a non-constant component in CMN makes the energy
resolution worse. 
For the 100~$\mu$m SSD, 
the signal and noise separation was improved by fitting 
CMN with a linear curve at a $\beta$-ray source test.
We conclude that a 100~$\mu$m SSD with analog VLSI readout 
can be used as a very thin tracking device in a future experiment. 
However, careful treatment of CMN is important for detecting MIP's.

\begin{ack}
This work has been supported by the Grant-in-Aid for General Science
Research (No. 07454057 and No. 09640358) by the Ministry of Education, 
Science and Culture.
\end{ack}

\end{document}